# Improved RIP Analysis of Orthogonal Matching Pursuit

Ray Maleh[a]

[a]L-3 Communications Mission Integration Division, Greenville, TX
10001 Jack Finney Blvd.
Greenville, TX 75402

Phone Number: 903-408-2605
Fax Number: 903-408-9036
Email: Ray.Maleh@L-3com.com

**Abstract:** Orthogonal Matching Pursuit (OMP) has long been considered a powerful heuristic for attacking compressive sensing problems; however, its theoretical development is, unfortunately, somewhat lacking. This paper presents an improved Restricted Isometry Property (RIP) based performance guarantee for $T$-sparse signal reconstruction that asymptotically approaches the conjectured lower bound given in Davenport et al. We also further extend the state-of-the-art by deriving reconstruction error bounds for the case of general non-sparse signals subjected to measurement noise. We then generalize our results to the case of K-fold Orthogonal Matching Pursuit (KOMP). We finish by presenting an empirical analysis suggesting that OMP and KOMP outperform other compressive sensing algorithms in average case scenarios. This turns out to be quite surprising since RIP analysis (i.e. worst case scenario) suggests that these matching pursuits should perform roughly T^0.5 times worse than convex optimization, CoSAMP, and Iterative Thresholding.

## 1. Introduction

During the last decade, Orthogonal Matching Pursuit (OMP) has become an important component of the toolbox of any mathematician or engineer working in the field of compressive sensing (CS). The algorithm originated as a statistical method for projecting multi-dimensional data onto interesting lower dimensional spaces [14]. It was then introduced to the sparse approximation world in its non-orthogonal form Matching Pursuit by Mallat et al. in [22]. Today, OMP is a highly celebrated algorithm with applications in medical imaging [18],[21], synthetic aperture radar (SAR) [2], wireless multi-path channel estimation [1], and others. For the unfamiliar reader, OMP is a greedy alternative to convex optimization (see [6] and [10]) that solves the under-determined linear equation:

$$y = \Phi x \tag{1.1}$$

where the vector $x$ is a sparse (or highly compressible) signal, the matrix $\Phi$ is a short, fat measurement matrix, and the vector $y$ is a small set of linear measurements of the signal. While being a powerful heuristic, OMP suffers from a lack of a decent theoretical analysis. Up until recently, only sparse approximation performance guarantees have been derived for OMP [15],[27]. These results depend on the coherence (or cumulative coherence) of the matrix $\Phi$. Furthermore, these results only bound the error $\|y - \tilde{y}\|_2$ where $\tilde{y}$ is an estimate, produced by OMP, which has a sparse representation in the column span of $\Phi$. In compressive sensing, we are primarily concerned with bounds on $\|x - \tilde{x}\|_2$, which do not trivially follow from bounds on $\|y - \tilde{y}\|_2$.

As stated earlier, OMP is an alternative to convex optimization that solves the basic CS problem $y = \Phi x$. While convex programs tend to be slow, their solutions enjoy powerful error bounds based on a restricted isometry property (RIP) [7]. Needell and Tropp [23] later showed that CoSAMP, an algorithm that is inherently similar to OMP, does possess similar RIP-based guarantees. The high-level reasoning for this is that, like convex programming, CoSAMP works globally by simultaneously trying to identify all the correct non-zero entries of a sparse vector $x$ at each iteration. On the other hand, OMP works locally by attempting to select one non-zero entry of $x$ per iteration. In [11], an RIP-based condition is derived that guarantees OMP's ability to recover a $T$-sparse vector. Also, a lower bound on the best possible RIP for OMP is suggested without proof. The main result of [11] is slightly tightened in [18]. The paper [19] offers an asymptotic improvement over [18], but this comes at the expense of additional coherency assumptions and the incorporation of large scaling constants. This work will expand and improve the results in [11] and [18] without the additional conditions imposed in [19]. In Section 3, we first derive a strong RIP-based result for strictly $T$-sparse signals that asymptotically approaches the lower bound conjectured in [11]. Then we deduce an error bound on $\|x - \tilde{x}\|_2$ that describes OMP's ability to estimate non-sparse signals in the presence of measurement noise. In Section 4, we generalize these results to the case of K-fold Orthogonal Matching Pursuit (KOMP) where $K$ entries of $x$ are recovered at every iteration. Section 5 features an empirical comparison of OMP, KOMP, and other popular CS algorithms. We will show that despite the fact that the RIP implies that OMP performs $\sqrt{T}$ times worse than convex optimization, OMP and KOMP still outperform other methods in average case scenarios.

## 2. Preliminaries

Throughout this paper, we will employ the following notational conventions. We let $x \in \mathbb{C}^N$ denote a signal of interest. We say that $x$ is $T$-sparse if it consists of at most $T$ non-zero components. More formally, we can write that $\|x\|_0 = T$ where $\|\cdot\|_0$ denotes the quasi-norm that counts the number of non-zero entries in its argument. We typically let $\Lambda$ denote the support of $x$. We define the $\ell_p$ norm of a signal $x$ as follows:

$$\|x\|_p = \left(\sum_{n=1}^{N} |x_n|^p\right)^{1/p} \quad 0 < p < \infty$$
$$\|x\|_\infty = \max_{1 \leq n \leq N} |x_n| \quad p = \infty \tag{2.1}$$

It is assumed that we do not have direct access to the signal $x$. Instead, we have access to a set of $M \ll N$ linear measurements of $x$ that take the form:

$$y = \Phi x \tag{2.2}$$

where $\Phi \in \mathbb{C}^{M \times N}$ is a measurement matrix and the signal $y \in \mathbb{C}^M$ contains the actual measurements. Our goal is to solve for $x$ given knowledge of only $\Phi$ and $y$. In practical applications, $\Phi$ is often either a random sub-Gaussian matrix or a sub-matrix of a discrete Fourier transform matrix [25]; however, in theory, $\Phi$ can be any matrix provided it satisfies a restricted isometry property, which is defined as follows:

**Definition 1:** A measurement matrix $\Phi$ satisfies a restricted isometry property (RIP) of order $T$ if there exists a constant $0 < \delta_T < 1$ such that

$$(1 - \delta_T)\|x\|_2^2 \leq \|\Phi x\|_2^2 \leq (1 + \delta_T)\|x\|_2^2 \tag{2.3}$$

for all signals $x$ that are $T$-sparse. The constant $\delta_T$ is called the restricted isometry number of order $T$.

An equivalent formulation of the RIP is that for every indexing set $\Lambda$ of size $T$, we have that the $M \times T$ sub-matrix $\Phi_\Lambda$ generated by selecting the $T$ columns of $\Phi$ corresponding to $\Lambda$, satisfies:

$$1 - \delta_T \leq \text{Eigenvalues}(\Phi_\Lambda^* \Phi_\Lambda) \leq 1 + \delta_T \tag{2.4}$$

It is immediately clear that for a given measurement matrix $\Phi$, the restricted isometry numbers $\delta_T$ form an increasing sequence. Furthermore, Needell et al. [23] show that the growth of these numbers must be sub-linear, i.e.

$$\delta_T \leq T\delta_2. \tag{2.5}$$

The kernel of the measurement matrix $\Phi$ induces the quotient space $\mathbb{C}^N/\ker(\Phi)$ which consists of the cosets $\langle x \rangle = \{u \in \mathbb{C}^N | \Phi x = \Phi u\}$. It is a straight-forward exercise to show that if $\delta_{2T} < 1$, then each coset can contain at most one $T$-sparse signal. Thus, the generally underdetermined system (2.2) is well-defined if $x$ is $T$-sparse and $\delta_{2T} < 1$.

Naïve methods for solving (2.2) involve slow combinatorial searches over all possible $M \times T$ submatrices of $\Phi$. This becomes an intractable problem for large $N$. Candes [7] showed that under the tighter restricted isometry condition $\delta_{2T} < \sqrt{2} - 1$, the system (2.2) can be solved via the convex optimization problem:

$$\hat{x} = \underset{u}{\operatorname{argmin}} \|u\|_1 \text{ subject to } \Phi u = y \tag{2.6}$$

If $x$ is not $T$-sparse or is corrupted by noise, then the constraint in (2.6) can be formulated as $\|\Phi u - y\|_2 < \epsilon$ for some parameter $\epsilon > 0$. In this case it can be shown that the estimate $\hat{x}$ satisfied an error bound of the form:

$$\|x - \hat{x}\|_2 \leq O(\|x - x_T\|_2) + O\left(\frac{\|x - x_T\|_1}{\sqrt{T}}\right) \tag{2.7}$$

The downside to the $\ell_1$ minimization problem shown in (2.6) is that it runs in polynomial time. There are faster algorithms that solve the same problem.

One classical greedy algorithm that solves the compressive sensing problem is Orthogonal Matching Pursuit (OMP). The OMP algorithm, which is also known as Forward Stepwise Regression in the data mining and statistical learning communities [17], is shown below.

| TABLE I |
|---|
| ORTHOGONAL MATCHING PURSUIT |
| **INPUTS:** Observations $y = \Phi x$<br>Measurement Matrix $\Phi$<br>Number of Iterations $T$ (typically equal to sparsity level) |
| **OUTPUTS:** $T$-sparse estimate $\tilde{x}$ of $x$.<br>Residual signal $r_T$.<br>Set of selected support indices $\Lambda_T$. |
| **PROCEDURE:**<br>-Initialize the residual $r_0 = y$ and indexing set $\Lambda_0 = \emptyset$.<br>-For $t$ from 1 to $T$<br>{<br>  -Find the column of $\Phi$ that maximizes the correlation with the residual $r_t$, i.e. let<br>$$\lambda_t = \underset{i}{\operatorname{argmax}} |\phi_i^* r_t|$$<br>  where $\phi_i$ denotes the $i$th column of $\Phi$.<br>  -Set $\Lambda_t = \Lambda_{t-1} \cup \{\lambda_t\}$.<br>  -Let $p_t$ denote the projection of $y$ onto the columns of $\Phi$ indexed by $\Lambda_t$, i.e.<br>$$p_t = \Phi_{\Lambda_t} \Phi_{\Lambda_t}^\dagger y.$$<br>  -Define the new residual as $r_t = y - p_t$.<br>}<br>-Generate the estimate $\tilde{x}$ of $x$ as follows:<br>$$\tilde{x}_i = \begin{cases} \Phi_{\Lambda_T}^\dagger y & i \in \Lambda_T \\ 0 & i \notin \Lambda_T \end{cases}.$$ |

A popular generalization of Orthogonal Matching Pursuit is K-fold Orthogonal Matching Pursuit (KOMP) [15]. This procedure is essentially identical to OMP except for the fact that $K$ columns of $\Phi$ are selected per iteration. Thus, at every step, KOMP increments its indexing set according to the rule $\Lambda_t = \Lambda_{t-1} \cup L_t$ where $L_t$ consists of the indices corresponding to the $K$ largest values of $|\phi_i^* r_t|$. Observe that OMP is a special case of KOMP when $K = 1$. While clearly faster than OMP, a surprising result of this paper is that KOMP can sometimes be more accurate than OMP as well.

Up until recently, the theoretical development of Orthogonal Matching Pursuit has been limited. In [16], a non-uniform "per signal" performance guarantee, which assumes a Gaussian random measurement ensemble, is derived. Other works, such as [15],[27], etc., develop results based on dictionary coherence and/or cumulative coherence. The seminal papers [3] and [23] demonstrate that the restricted isometry property can be used in the analysis of the related greedy algorithms Iterative Thresholding and CoSAMP (Compressive Sensing Matching Pursuit). These works motivated the theoretical thrust to determine whether OMP enjoys an RIP-based performance guarantee as well. In [11], Davenport et al. demonstrate that OMP can successfully recover any $T$-sparse signal provided that the measurement matrix satisfies the RIP:

$$\delta_T < \frac{1}{3\sqrt{T}}. \tag{2.8}$$

The authors further allude to the unachievable lower bound $\delta_T \sim 1/\sqrt{T}$. In [18], Liu et al. improve this result slightly and obtain:

$$\delta_{T+1} < \frac{1}{(1+\sqrt{2})\sqrt{T}}. \tag{2.9}$$

Along these same lines, Lipshitz [19] shows that with additional stringent dictionary coherence constraints, the result can be asymptotically improved to:

$$\delta_{CT^{1.2}} < \frac{c}{T^{0.2}}. \tag{2.10}$$

for some very large constant $C \sim 2 \times 10^5$ and some very small constant $c \sim 10^{-6}$. Unfortunately, because of the magnitudes of these constants, the benefit of this result will be very difficult to realize in practical compressive sensing problems.

In Section 3, we improve upon the results (2.8) and (2.9) and show, without additional coherence assumptions, that OMP can recover any $T$-sparse signal provided

$$\delta_{T+1} < \frac{1}{1+\sqrt{T}}, \tag{2.11}$$

which is a highly near-optimal figure. In addition, we show that for a signal $x$ that is not $T$-sparse and/or has measurements that are corrupted by noise, OMP will obtain an estimate $\tilde{x}$ of $x$ satisfying:

$$\|x - \tilde{x}\|_2 \leq O(\sqrt{T})\|x - x_T\|_2 + O(\|x - x_T\|_1) + O(\sqrt{T})\|w\|_2 \tag{2.12}$$

where $w$ represents the measurement noise and $x_T$ is the optimal $T$-sparse estimate of $x$, i.e. $x$ truncated to its $T$ strongest entries. In Section 4, we extend these results to the case of KOMP and compare the theoretical performance of OMP and KOMP. In particular, we illustrate situations in which KOMP performs better than OMP. In Section 5, we augment this discussion by using empirical methods to compare OMP, KOMP, and the various other popular CS algorithms of the day.

## 3. Regular Orthogonal Matching Pursuit

We begin by citing two lemmas that will be used repeatedly throughout this analysis:

**Lemma 1:** Let $x$ be a $T$-sparse signal with support set $\Lambda$. Let $A$ be any subset of $\{1, 2, \cdots, N\}$ such that $A \cap \Lambda = \emptyset$. Let $\Phi$ be a measurement matrix with restricted isometry numbers $\delta_T$. Then the following two properties are true:

$$\|\Phi_\Lambda^* \Phi x\|_2 \geq (1 - \delta_T) \|x\|_2 \tag{3.1}$$

and

$$\|\Phi_A^* \Phi x\|_2 \leq \delta_{T+|A|} \|x\|_2. \tag{3.2}$$

**Lemma 2:** Let $\Phi$ be a measurement matrix with restricted isometry number $\delta_T$. Let $x$ be any signal. Then

$$\|\Phi x\|_2 \leq \sqrt{1 + \delta_T} \left( \|x\|_2 + \frac{1}{\sqrt{T}} \|x\|_1 \right). \tag{3.3}$$

Both lemmas are proved in [23]. The first lemma is an immediate consequence of the fact that $\Phi$ is nearly unitary with respect to $T$-sparse signals. The second lemma extends the restricted isometry energy bounds to non-sparse signals. With these lemmas in mind, we are now prepared to develop a sufficient condition under which Orthogonal Matching Pursuit will recover any $T$-sparse signal from its measurements:

**Theorem 1:** Suppose that $\Phi$ is a measurement matrix whose RIP constant satisfies

$$\delta_{T+1} < \frac{1}{1 + \sqrt{T}}. \tag{3.4}$$

Then, OMP will recover any $T$-sparse signal from its measurements $\Phi x$.

*Proof.* Let $x$ be any $T$-sparse signal with support $\Lambda$. At iteration $t$, suppose that OMP has only selected correct atoms. Let $r_t$ be the current residual. Then $r_t = \Phi c_t$ where $c_t$ is also supported on $\Lambda$. Now observe that, by Lemma 1 and the fact that the RIP constants are increasing in $T$,

$$\|\Phi_\Lambda^* \Phi x\|_2 \geq (1 - \delta_T) \|c_t\|_2 \geq (1 - \delta_{T+1}) \|c_t\|_2. \tag{3.5}$$

This implies that

$$\|\Phi_\Lambda^* \Phi c_t\|_\infty \geq \frac{(1 - \delta_{T+1})}{\sqrt{T}} \|c_t\|_2. \tag{3.6}$$

Now let $i$ be any element not in $\Lambda$. Then observe that

$$\left\|\Phi_{\{i\}}^* \Phi c_t\right\|_\infty = \left\|\Phi_{\{i\}}^* \Phi c_t\right\|_2 \leq \delta_{T+1} \|c_t\|_2. \tag{3.7}$$

This implies that

$$\left\|\Phi_{\Lambda^c}^* \Phi c_t\right\|_\infty = \max_{i \notin \Lambda} \left\|\Phi_{\{i\}}^* \Phi c_t\right\|_\infty \leq \delta_{T+1} \|c_t\|_2. \tag{3.8}$$

OMP will recover the next atom correctly if

$$\|\Phi_\Lambda^* \Phi c_t\|_\infty \geq \left\|\Phi_{\Lambda^c}^* \Phi c_t\right\|_\infty \tag{3.9}$$

which will happen if

$$\frac{(1-\delta_{T+1})}{\sqrt{T}}\|c_t\|_2 \geq \delta_{T+1}\|c_t\|_2. \tag{3.10}$$

This is equivalent to (3.4). □

This sufficient condition is significantly stronger than the one demonstrated in [11]. In fact, it is very close to the unachievable bound $1/\sqrt{T}$ suggested in the same work. Thus, our result is highly near optimal.

We next extend this result to more general signals. Certainly, if $x$ is not $T$-sparse, then it is impossible for OMP to recover $x$ perfectly in $T$ iterations. However, we are interested in determining how close OMP's $T$-term approximation error is to the optimal $T$-term representation $x_T$ of $x$. The following theorem answers this precise question:

**Theorem 2:** Let $\Phi$ be a measurement matrix that satisfies the RIP shown in (3.4). Let $x \in \mathbb{C}^N$ be any signal with optimal $T$-term approximation $x_T$. Let $\Lambda = \text{supp}(x_T)$ and let $x_{T^c} = x - x_T$. Suppose OMP has noisy measurements of the form $y = \Phi x + w = \Phi x_T + e$ where $e = \Phi x_{T^c} + w$. Then, after $T$ iterations, OMP will recover an estimate $\tilde{x}$ of $x$ that satisfies:

$$\|x - \tilde{x}\|_2 \leq \left(1 + C_1(T)\right)\|x - x_T\|_2 + \frac{C_1(T)}{\sqrt{T}}\|x - x_T\|_1 + C_1(T)\|w\|_2 \tag{3.11}$$

where, for reasonable RIP numbers, $C_1(T)$ grows asymptotically like $\sqrt{T}$.

*Proof.* First suppose that at iteration $t$, OMP has selected only atoms indexed in $\Lambda$. At iteration $t + 1$, OMP will select another atom from $\Lambda$ provided the greedy selection condition

$$\frac{\|\Phi_{\Lambda^c}^* r_t\|_\infty}{\|\Phi_\Lambda^* r_t\|_\infty} < 1 \tag{3.12}$$

is satisfied. Now rewrite the residual as $r_t = \Phi_\Lambda(x_T - a_t) + e$. Here, $a_t$ is the coefficient vector of the projection of $y$ onto the currently selected atoms. Then one can bound the numerator of (3.12) by:

$$\begin{aligned}\|\Phi_{\Lambda^c}^* r_t\|_\infty &\leq \|\Phi_{\Lambda^c}^* \Phi_\Lambda(x_T - a_t) + \Phi_{\Lambda^c}^* e\|_\infty & (3.13)\\ &\leq \|\Phi_{\Lambda^c}^* \Phi_\Lambda(x_T - a_t)\|_\infty + \|\Phi_{\Lambda^c}^* e\|_\infty & (3.14)\\ &\leq \delta_{T+1}\|(x_T - a_t)\|_2 + \|e\|_2 & (3.15)\end{aligned}$$

On the other hand, the denominator can be bounded from below by:

$$\begin{aligned}\|\Phi_\Lambda^* r_t\|_\infty &\geq \frac{1}{\sqrt{T}}\|\Phi_\Lambda^* r_t\|_2 & (3.16)\\ &= \frac{1}{\sqrt{T}}\|\Phi_\Lambda^* \Phi_\Lambda(x_T - a_t) + \Phi_\Lambda^* e\|_2 & (3.17)\\ &\geq \frac{\|\Phi_\Lambda^* \Phi_\Lambda(x_T - a_t)\|_2 - \|\Phi_\Lambda^* e\|_2}{\sqrt{T}} & (3.18)\\ &\geq \frac{(1-\delta_{T+1})}{\sqrt{T}}\|(x_T - a_t)\|_2 - \sqrt{\frac{1+\delta_{T+1}}{T}}\|e\|_2. & (3.19)\end{aligned}$$

A sufficient condition for the next atom to be selected from $\Lambda$ is that the numerator is less than the denominator. This is guaranteed if

$$\delta_{T+1}\|(x_T - a_t)\|_2 + \|e\|_2 < \frac{(1-\delta_{T+1})}{\sqrt{T}}\|(x_T - a_t)\|_2 - \sqrt{\frac{1+\delta_{T+1}}{T}}\|e\|_2. \qquad (3.20)$$

We can rearrange terms to obtain:

$$\|x_T - a_t\|_2 > \frac{\sqrt{T} + \sqrt{1+\delta_{T+1}}}{1 - \delta_{T+1}(1+\sqrt{T})}\|e\|_2. \qquad (3.21)$$

Now let $t^*$ denote the first iteration where (3.21) does not hold. By definition of OMP, $\tilde{x} = a_T$. We have:

$$\|x - \tilde{x}\|_2 \leq \|x_T - \tilde{x}\|_2 + \|x_{T^c}\|_2 \qquad (3.22)$$

$$\leq \frac{1}{\sqrt{1-\delta_{2T}}}\|\Phi_{\Lambda'}(x_T - \tilde{x})\|_2 + \|x_{T^c}\|_2 \qquad (3.23)$$

where $\Lambda' = \Lambda \cup \text{supp}(\tilde{x})$ which has cardinality at most $2T$. It is possible to further bound the left hand side by:

$$\|x - \tilde{x}\|_2 \leq \frac{\|\Phi_{\Lambda'}(x_T - \tilde{x}) + e\|_2 + \|e\|_2}{\sqrt{1-\delta_{2T}}} + \|x_{T^c}\|_2 \qquad (3.24)$$

$$\leq \frac{\|\Phi_{\Lambda'}(x_T - a_{t^*}) + e\|_2 + \|e\|_2}{\sqrt{1-\delta_{2T}}} + \|x_{T^c}\|_2 \qquad (3.25)$$

$$\leq \frac{\|\Phi_{\Lambda'}(x_T - a_{t^*})\|_2 + 2\|e\|_2}{\sqrt{1-\delta_{2T}}} + \|x_{T^c}\|_2 \qquad (3.26)$$

$$\leq \frac{\sqrt{1+\delta_T}}{\sqrt{1-\delta_{2T}}}\|x_T - a_{t^*}\|_2 + \frac{2\|e\|_2}{\sqrt{1-\delta_{2T}}} + \|x_{T^c}\|_2 \qquad (3.27)$$

where the second inequality comes from the fact that in OMP, the residual is always decreasing in magnitude regardless whether the selected atoms are from $\Lambda$ or not. Since (3.21) does not hold for $t^*$, it follows that:

$$\|x - \tilde{x}\|_2 \leq C_1'(T)\|e\|_2 + \|x_{T^c}\|_2 \qquad (3.28)$$

where

$$C_1'(T) = \left(\frac{\sqrt{1+\delta_T}}{\sqrt{1-\delta_{2T}}}\right)\left(\frac{\sqrt{T} + \sqrt{1+\delta_{T+1}}}{1 - \delta_{T+1}(1+\sqrt{T})}\right) + \frac{2}{\sqrt{1-\delta_{2T}}} \qquad (3.29)$$

We further bound $\|e\|_2$ by:

$$\|e\|_2 = \|\Phi x_{T^c} + w\|_2 \qquad (3.30)$$

$$\leq \|\Phi x_{T^c}\|_2 + \|w\|_2 \qquad (3.31)$$

$$\leq \sqrt{1+\delta_T}\left(\|x_{T^c}\|_2 + \frac{1}{\sqrt{T}}\|x_{T^c}\|_1\right) + \|w\|_2 \qquad (3.32)$$

Finally, let $C_1(T) = \sqrt{1+\delta_T}C_1'(T)$ to obtain that

$$\|x - \tilde{x}\|_2 \leq (1+C_1)\|x_{T^c}\|_2 + \frac{C_1\|x_{T^c}\|_1}{\sqrt{T}} + C_1\|w\|_2 \qquad (3.33)$$

$$= \left(1 + C_1(T)\right)\|x - x_T\|_2 + \frac{C_1(T)\|x - x_T\|_1}{\sqrt{T}} + C_1(T)\|w\|_2 \qquad (3.34)$$

as was to be shown. □

The fact that $C_1(T)$ grows asymptotically like $\sqrt{T}$ should be no surprise: it is already well known [27] that, under mild coherence constraints, the signal observations obey the bound:

$$\|y - \tilde{y}\|_2 \leq O(\sqrt{T})\|y - y_T\|_2 \qquad (3.35)$$

where $y = \Phi x$, $\tilde{y} = \Phi \tilde{x}$, and $y_T$ is the optimal $T$-term representation of $y$ using the columns of $\Phi$. We note that it is not necessarily the case that $y_T = \Phi x_T$.

The novelty of our result is that we have derived an error bound on the signal itself, not simply on its measurements. In a sparse approximation problem where $y$ is the signal that we're trying to estimate using the columns of $\Phi$, then a bound such as (3.35) is sufficient. However, in compressive sensing, if $\|y - \tilde{y}\|_2 = 0$, then all that we may conclude is that $x = \tilde{x} + x_0$ where $x_0 \in \ker(\Phi)$. If the sparsity $T$ is not sufficiently small, then $x_0$ may be a non-zero vector and our estimate may be quite inaccurate. Thus, we claim that the result in Theorem 2 is more powerful than the sparse approximation results of the past.

## 4. K-fold Orthogonal Matching Pursuit

A popular extension of Orthogonal Matching Pursuit is K-fold Orthogonal Matching Pursuit (KOMP). KOMP is almost identical to OMP except for the fact that $K$ atoms are selected per iteration instead of 1. KOMP has two main advantages over OMP which are somewhat mutually exclusive. The first one is speed: Given a $T$-sparse signal, one may use KOMP to recover the signal in $T/K$ iterations versus the usual $T$ iterations. This yields a significant reduction in run-time especially since the number of least-squares projections has been cut down by a factor of $K$. Unfortunately, for accuracy, this method requires that all $K$ atoms selected per iteration be correct. Very few measurement matrices enjoy enough coherence to allow for the correct selection of so many atoms without some sort of re-orthogonalization. As a result, we choose to exploit the second mutually exclusive advantage of KOMP over OMP: Running $T$ iterations of KOMP will select a set $S'$ of $KT$ indices where, with good probability, our signal's support set $S$ will be contained in $S'$. Thus, we effectively use KOMP to narrow down all the possible signal indices to the top $KT$ candidates. Then, assuming $K$ is not too large, we can perform a least-squares projection of our measurements onto the span of the selected $KT$ columns of $\Phi$ in order to identify the exact support set and recover the signal. All of this can be done in runtime commensurate to that achievable with $T$ iterations of OMP.

To analyze the performance of KOMP, we first need to define the top-K norm:

**Definition 2:** Let $x$ be a signal with sorted entries $x_{(1)}, x_{(2)}, \cdots, x_{(N)}$ where $|x_{(1)}| \geq |x_{(2)}| \geq \cdots \geq |x_{(N)}|$. Then the top-K norm is defined to be:

$$\|x\|_{\text{Top}K} = \left\|(x_{(1)}, x_{(2)}, \cdots, x_{(N)})\right\|_2 = \sqrt{\sum_{k=1}^{K} |x_{(k)}|^2}. \qquad (4.1)$$

In essence, the top-K norm of a vector is the $\ell_2$ norm of its top $K$ entries. It is not difficult to show that this is a well-defined norm on $\mathbb{C}^N$.

We begin by examining KOMP's ability to correctly select all the correct non-zero entries of a $T$-sparse signal in addition to at most $(K - 1)T$ incorrect entries:

**Theorem 3:** Let $\Phi$ be a measurement matrix satisfying

$$\delta_{T+(K-2)t+K} < \frac{1}{1+\sqrt{\frac{T-t+1}{K}}} \quad (4.2)$$

for each iteration $t = 1, \cdots, T$. Assuming also that $\delta_{KT} < 1$, then KOMP will recover any $T$-sparse signal $x$ from its measurements.

*Proof.* Observe first that KOMP will select all the correct entries in $T$ iterations if it selects at least one correct entry per iteration. Assume that after $t$ iterations, KOMP has selected at least $t$ correct indices specified by the set $A_t$ and no more than $(K-1)t$ incorrect indices specified by the set $B_t$. We define the set $\Lambda_t = (\Lambda \cup B_t) \setminus A_t$, which has no more than $T + (K-2)t$ elements. Defining $c_t$ as in Theorem 1, we see that $c_t \in \text{colspan}(\Phi_{\Lambda_t})$. As a result, a sufficient condition to ensure that KOMP selects at least one correct index at iteration $t+1$ is that

$$\left\| \Phi_{\Lambda_t^c}^* \Phi c_t \right\|_{\text{TopK}} < \left\| \Phi_{\Lambda_t}^* \Phi c_t \right\|_{\text{TopK}}. \quad (4.3)$$

We can find an upper bound on the left hand side as follows:

$$\left\| \Phi_{\Lambda_t^c}^* \Phi c_t \right\|_{\text{TopK}} = \max_{\substack{B \subseteq \Lambda_t^c \\ |B|=K}} \left\| \Phi_B^* \Phi c_t \right\|_2 \quad (4.4)$$

$$\leq \delta_{T+(K-2)t+K} \|c_t\|_2. \quad (4.5)$$

We next calculate a lower bound on the right hand side:

$$\left\| \Phi_{\Lambda_t}^* \Phi c_t \right\|_{\text{TopK}} \geq \frac{\left\| \Phi_{\Lambda_t}^* \Phi c_t \right\|_2}{\sqrt{(T-t+1)/K}} \quad (4.6)$$

$$\geq \frac{\delta_{T+(K-2)t} \|c_t\|_2}{\sqrt{(T-t+1)/K}} \quad (4.7)$$

$$\geq \frac{\delta_{T+(K-2)t+K} \|c_t\|_2}{\sqrt{(T-t+1)/K}} \quad (4.8)$$

From these bounds, one can use simple algebra to show that (4.2) is a sufficient condition for (4.3). Since $\delta_{KT} < 1$, a simple least squares projection onto the selected columns of $\Phi$ will recover the desired signal exactly. □

For the special case $K = 2$, (4.2) takes the form:

$$\delta_{T+2} < \frac{1}{1+\sqrt{T/2}}. \quad (4.9)$$

For relatively large $T$, $\delta_{T+2} \approx \delta_{T+1}$ and, therefore, we see that 2-OMP enjoys a significantly stronger performance guarantee than regular OMP. This should make intuitive sense since we are allowing for the selection of up to $T$ incorrect atoms. As long as $\delta_{2T} < 1$ (which is guaranteed by (4.9) for $T > 1$), the least squares projection of the measurements $y = \Phi x$ onto the space of selected columns will yield zeros for all incorrectly chosen entries. In general, the performance of KOMP will improve for increasing $K$ until the final least squares projection becomes unstable. At this point, performance will degrade rapidly.

We can extend the KOMP result to general signals via the following Theorem:

**Theorem 4:** Let $\Phi$ be a measurement matrix that satisfies the RIP shown in (4.2). Let $x \in \mathbb{C}^N$ be any signal with optimal $T$-term approximation $x_T$. Let $\Lambda = \text{supp}(x_T)$ and let $x_{T^c} = x - x_T$. Suppose OMP has noisy measurements of the form $y = \Phi x + w = \Phi x_T + e$ where $e = \Phi x_{T^c} + w$. Then, after $T$ iterations, KOMP will recover a $KT$-sparse estimate $\tilde{x}$ of $x$ that satisfies:

$$\|x - \tilde{x}\|_2 \leq \big(1 + C_K(T)\big)\|x - x_T\|_2 + \frac{C_K(T)}{\sqrt{T}}\|x - x_T\|_1 + C_K(T)\|w\|_2 \tag{4.10}$$

where, for reasonable RIP numbers, $C_K(T)$ grows asymptotically like $\sqrt{T/K}$.

*Proof.* The argument is very similar in nature to that in Theorem 2. We retain the notation used in the proofs of Theorem 2 and Theorem 3. As before, we suppose that at iteration $t$, OMP has selected at least one atom indexed by $\Lambda$ per iteration. At iteration $t+1$, OMP will select at least one atom from $\Lambda$ provided the greedy selection condition

$$\frac{\|\Phi^*_{\Lambda_t} r_t\|_{\text{Top}K}}{\|\Phi^*_{\Lambda^c_t} r_t\|_{\text{Top}K}} < 1 \tag{4.11}$$

is satisfied where $\Lambda_t = (\Lambda \cup B_t) \setminus A_t$ has no more than $T + (K-2)t$ elements. Now rewrite the residual as $r_t = \Phi_{\Lambda_t}(x_T - a_t) + e$. We bound the numerator from below as follows:

$$\|\Phi^*_{\Lambda^c_t} r_t\|_{\text{Top}K} = \max_{\substack{B \subseteq \Lambda^c_t \\ |B|=K}} \|\Phi^*_B(\Phi_{\Lambda_t}(x_T - a_t) + e)\|_2 \tag{4.12}$$

$$\leq \max_{\substack{B \subseteq \Lambda^c_t \\ |B|=K}} \|\Phi^*_B \Phi_{\Lambda_t}(x_T - a_t)\|_2 + \max_{\substack{B \subseteq \Lambda^c_t \\ |B|=K}} \|\Phi^*_B e\|_2 \tag{4.13}$$

$$\leq \delta_{T_{t,K}} \|x_T - a_t\|_2 + \sqrt{1 + \delta_K} \|e\|_2 \tag{4.14}$$

where $T_{t,K} = T + (K-2)t + K$. Next, we derive a lower bound for the denominator:

$$\|\Phi^*_{\Lambda_t} r_t\|_{\text{Top}K} \geq \frac{\|\Phi^*_{\Lambda_t}(\Phi_{\Lambda_t}(x_T - a_t) + e)\|_2}{\sqrt{(T-t+1)/K}} \tag{4.15}$$

$$\geq \frac{\|\Phi^*_{\Lambda_t} \Phi_{\Lambda_t}(x_T - a_t)\|_2 - \|\Phi^*_{\Lambda_t} e\|_2}{\sqrt{(T-t+1)/K}} \tag{4.16}$$

$$\geq \frac{\delta_{T_{t,K}} \|x_T - a_t\|_2 - \sqrt{1 + \delta_{T_{t,K}}} \|e\|_2}{\sqrt{(T-t+1)/K}} \tag{4.17}$$

Our bounds imply that a sufficient condition for (4.11) is that

$$\|x_T - a_t\|_2 > \frac{\sqrt{(T-t+1)/K}\sqrt{1 + \delta_K} + \sqrt{1 + \delta_{T_{t,K}}}}{1 - \delta_{T_{t,K}}(1 + \sqrt{(T-t+1)/K})} \|e\|_2. \tag{4.18}$$

Now let $t^*$ denote the first iteration where this bound does not hold. By definition of KOMP, $\tilde{x} = a_T$. We have:

$$\|x - \tilde{x}\|_2 = \|x_T - \tilde{x} + x_{T^c}\|_2 \leq \|x_T - \tilde{x}\|_2 + \|x_{T^c}\|_2 \tag{4.19}$$

$$\leq \frac{1}{\sqrt{1 - \delta_{(K+1)T}}} \|\Phi_{\Lambda'}(x_T - \tilde{x})\|_2 + \|x_{T^c}\|_2 \tag{4.20}$$

where $\Lambda' = \Lambda \cup \text{supp}(\tilde{x})$ which has cardinality at most $(K+1)T$. It is possible to further bound the left hand side by:

$$\|x - \tilde{x}\|_2 \leq \frac{\|\Phi_{\Lambda'}(x_T - \tilde{x}) + e\|_2 + \|e\|_2}{\sqrt{1 - \delta_{(K+1)T}}} + \|x_{T^c}\|_2 \tag{4.21}$$

$$\leq \frac{\|\Phi_{\Lambda'}(x_T - a_{t^*}) + e\|_2 + \|e\|_2}{\sqrt{1 - \delta_{(K+1)T}}} + \|x_{T^c}\|_2 \tag{4.22}$$

$$\leq \frac{\|\Phi_{\Lambda'}(x_T - a_{t^*})\|_2 + 2\|e\|_2}{\sqrt{1 - \delta_{(K+1)T}}} + \|x_{T^c}\|_2 \tag{4.23}$$

$$\leq \frac{\sqrt{1 + \delta_{T_{t^*,K}}}}{\sqrt{1 - \delta_{(K+1)T}}} \|x_T - a_{t^*}\|_2 + \frac{2\|e\|_2}{\sqrt{1 - \delta_{(K+1)T}}} + \|x_{T^c}\|_2 \tag{4.24}$$

$$\leq \frac{\sqrt{1 + \delta_{T_K}}}{\sqrt{1 - \delta_{(K+1)T}}} \|x_T - a_{t^*}\|_2 + \frac{2\|e\|_2}{\sqrt{1 - \delta_{(K+1)T}}} + \|x_{T^c}\|_2 \tag{4.25}$$

where $T_K = T + (K-2)T + K$. The second inequality comes from the fact that in OMP, the residual is always decreasing in magnitude regardless of which atoms are selected. Now let

$$C_K''(T) = \frac{\sqrt{T/K}\sqrt{1 + \delta_K} + \sqrt{1 + \delta_{T_K}}}{1 - \delta_{T_K}(1 + \sqrt{T/K})}. \tag{4.26}$$

Since $\|x_T - a_{t^*}\|_2 \leq C_K''(T)\|e\|_2$, which follows from (4.18), we have that

$$\|x - \tilde{x}\|_2 \leq C_K'(T)\|e\|_2 + \|x_{T^c}\|_2 \tag{4.27}$$

where

$$C_K'(T) = \frac{\sqrt{1 + \delta_{T_K}}}{\sqrt{1 - \delta_{(K+1)T}}} C_K''(T) + \frac{2}{\sqrt{1 - \delta_{(K+1)T}}} \tag{4.28}$$

We use our previous bound

$$\|e\|_2 \leq \sqrt{1 + \delta_T} \left( \|x_{T^c}\|_2 + \frac{1}{\sqrt{T}} \|x_{T^c}\|_1 \right) + \|w\|_2 \tag{4.29}$$

and the definition $C_K(T) = \sqrt{1 + \delta_T} C_K'(T)$ to obtain:

$$\|x - \tilde{x}\|_2 \leq (1 + C_K)\|x_{T^c}\|_2 + \frac{C_K \|x_{T^c}\|_1}{\sqrt{T}} + C_K \|w\|_2 \tag{4.30}$$

$$\leq (1 + C_K(T))\|x - x_T\|_2 + \frac{C_K(T)\|x - x_T\|_1}{\sqrt{T}} + C_K(T)\|w\|_2 \tag{4.31}$$

as was to be shown. □

We observe that the constants $C_K(T)$ form a decreasing sequence with respect to $K$, which suggests that the errors $\|x - \tilde{x}\|_2$ decrease as we let $K$ increase. Of course, one may argue that since for each $K$, $\tilde{x}$ is $KT$-sparse, and therefore, it is unfair to compare reconstructions using different values of $K$. As a result, we will let $\tilde{x}_T$ denote the truncation of $\tilde{x}$ to its top $T$ values. It is fairly straight-forward to show the bound

$$\|x - \tilde{x}_T\|_2 \leq 2\|x - \tilde{x}\|_2 + \|x - x_T\|_2. \tag{4.32}$$

which implies the following corollary.

**Corollary 1:** Let $\Phi$ be a measurement matrix that satisfies the RIP shown in (4.2). Then, for any signal $x$, KOMP will return a $KT$-sparse estimate $\tilde{x}$ whose $T$-sparse truncation $\tilde{x}_T$ satisfies:

$$\|x - \tilde{x}_T\|_2 \leq \left(3 + 2C_K(T)\right)\|x - x_T\|_2 + \frac{C_K(T)}{\sqrt{T}}\|x - x_T\|_1 + 2\|w\|_2 \tag{4.33}$$

We can now make a comparison of OMP and KOMP by comparing the constants $C_1(T)$ against $2C_K(T) + 2$. Assume for the moment that the restricted isometry numbers obey $\delta_\ell = \delta_2 \ell^\beta$ for some $\beta \leq 1$. For sparsity level $T = 100$, $\beta \in \{.3, .8, .95\}$, and $\delta_2 = .00015$, we calculated the above constants and plotted them in Figure 1.

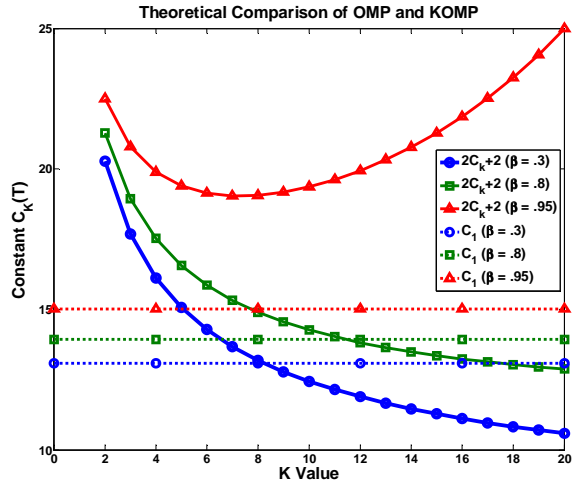

*Figure 1: Comparison of OMP and KOMP constants $C_1(T)$ and $2C_K(T) + 2$.*

In the case of $\beta = .3$ and $\beta = .8$, we see that KOMP achieves better results than OMP when $K \geq 9$ and $K \geq 12$ respectively. Eventually, when the RIP constants for sparsity level $100K$ become too large (as in the case $\beta = .95$), the constant $C_K(T)$ begins to increase rapidly. In this latter case, KOMP does not achieve a stronger error bound than OMP regardless of the choice of $K$. As we can see, selecting an appropriate $K$ can be challenging. If $K$ is selected too small, then KOMP's performance will be suboptimal when compared against OMP and KOMP with larger $K$. However, if $K$ is selected too large, then instability may arise due to the fact that the underlying RIP constants are becoming increasingly large as well. Selecting the right value of $K$ is, thus, somewhat of an art form: Intuition derived from copious experimentation is extremely helpful.

## 5. Experimental Results

An observation that one will quickly make regarding compressive sensing algorithms is that, in practice, they all work better than predicted by their respective theoretical guarantees. In other words, the restricted isometry property only affords relatively weak sufficient conditions specifying when some algorithm can exactly recover any signal with a given number $T$ of non-zero entries. The reason for this is that RIPs provide worse case estimates that may not appear often in practice. In order to address this issue, much work has been done in performing "average-case" analyses on compressive sensing algorithms (see [26], [12], etc.). In these works, theoretical results are obtained regarding the various algorithms' performance in recovering commonplace sparse signals, e.g. with Gaussian or binary coefficients. For our purposes, we will empirically perform a similar analysis by designing several experiments which are shown below.

In the first experiment, for every sparsity level $T$ from 4 to 52 in increments of 4, the following test was repeated 100 times: A $T$-sparse Gaussian signal of length 256 was generated and measurements of the form $\Phi x$ were collected where $\Phi$ is a $100 \times 256$ Gaussian random matrix (selected differently each time). Then the following algorithms were used to recover $x$: OMP, 2-OMP, Hybrid 0.2-OMP[1], $CoSAMP_2$ [22], $CoSAMP_1$[2], Iterative Thresholding [3],[8],[13], and Basis Pursuit [5],[7],[9]. Both versions of CoSAMP were run with 10 iterations. For Iterative Thresholding, the Hard Thresholding routine in the Sparsify MATLAB package [4] was used with all parameters being selected optimally by the software. We used the L1-Magic package [24] for Basis Pursuit with the default settings. The two performance criteria evaluated were the probability of exact reconstruction (within a 1% tolerance for relative error) and the runtime. Plots of the results are shown below in Figure 2 and Figure 3. In terms of exact reconstruction probability, Basis Pursuit did slightly better than OMP. However, the modifications proposed in Section 2.2 came in quite handy because 2-OMP and Hybrid 0.2 OMP both outperformed Basis Pursuit. Thus, the suggestion of allowing multiple atoms to be selected per iteration was exactly what was needed to give OMP the extra boost to put it on top. $CoSAMP_1$ performed better than $CoSAMP_2$ and Iterative Thresholding fell roughly in between in this particular experimental setup. With respect to runtime, all of the algorithms were very fast with the exception of convex optimization. These algorithms took no more than a tenth of a second to run whereas $\ell_1$ minimization took about a half of a second. The overall conclusion of this experiment is that 2-OMP was the best overall performer.

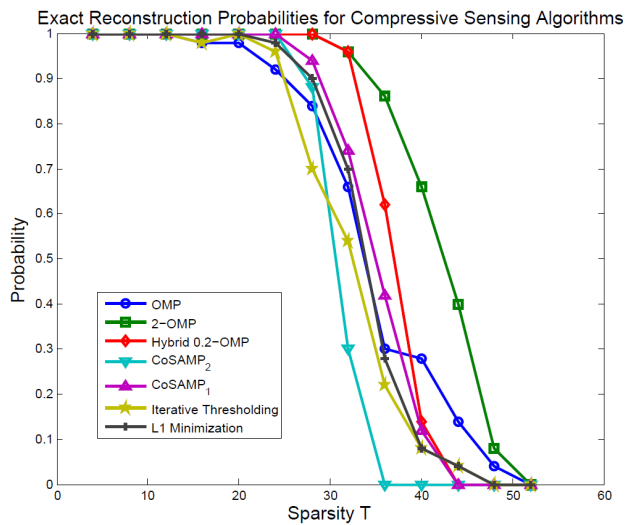

*Figure 2: Probability of exact reconstruction of T-sparse signals using various compressive sensing algorithms.*

---

[1] Hybrid $\alpha$-OMP is variation of KOMP where at iteration $t$, the top $\alpha(T - t + 1)$ atoms are selected. Thus, it selects more atoms during earlier iterations and fewer atoms in subsequent iterations.
[2] $CoSAMP_1$ is a variation of regular $CoSAMP_2$ (see [22]) where $T$ atoms are selected per iteration as opposed to the standard $2T$.

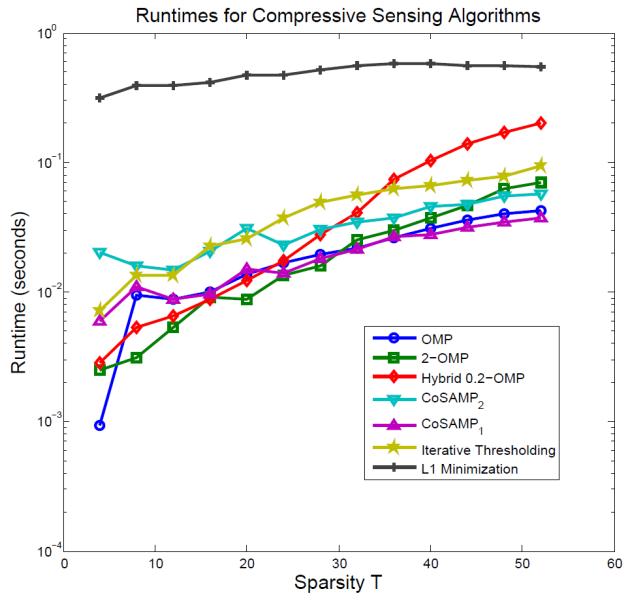

*Figure 3: Runtimes of various compressive sensing algorithms when recovering T-sparse signals.*

Of course, the above experiment only compares the various compressive sensing algorithms with respect to their abilities to recover sparse signals. In the next experiment, the objective signals were not allowed to strictly be sparse. Here, 20 instances of a signals of length 256 were generated with exponentially decaying coefficients in random locations. The decay rate was given by $|x_{(n)}| \leq 0.9^n$. The signals were reconstructed using the same algorithms and sparsity parameters varying from $T = 4$ to $T = 52$ in increments of four. Figure 4 shows the various average $\ell_2$ reconstruction errors produced by these algorithms.

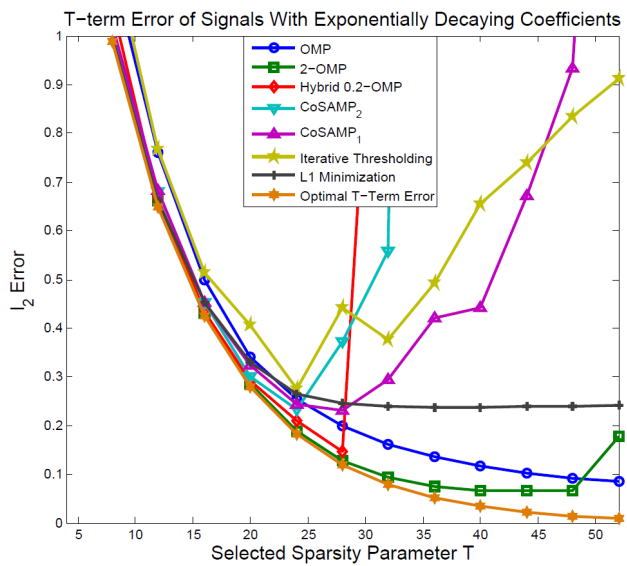

*Figure 4: Average T-term reconstruction errors in recovering signals with exponentially decaying coefficients generated by the various compressive sensing algorithms as a function of the sparsity parameter T.*

In this experiment, OMP and its variants outperformed the other algorithms. In fact, the $T$-term error produced by 2-OMP is nearly identical to the optimal $T$-term error up until around $T = 25$. The L1-minimization error converges to around 0.24 whereas the true optimal error should converge to zero. An interesting point to note is that all of the above greedy algorithms ultimately experience a sudden and significant breakdown in performance when $T$ is taken too large. This is because of the instability that arises from computing projections when the underlying restricted isometry numbers approach unity. In other words, the more vectors that are being processed at any particular iteration, the greater the instability. This makes algorithms such as Iterative Thresholding and CoSAMP, which process a large set of atoms right from the start, highly susceptible to breakdown if care is not selected in choosing an appropriate sparsity level $T$. In these cases, $T$ becomes a highly sensitive parameter that can corrupt the output very suddenly and swiftly. On the other hand, OMP and its variants are more robust with respect to tolerating a large value of $T$. This is because these algorithms select no more than a few atoms per iteration. Thus, any instability that may result from a poor choice of $T$ will defer itself to later iterations. The first several selected atoms will remain correct. As a result, if one observes instability beginning to develop in the matching pursuit, then he/she can backtrack a few iterations and simply decide to stop there. This is not an option with Iterative Thresholding and CoSAMP. Ultimately, all of the greedy algorithms will experience a breakdown in performance; however, OMP and its variants are structured so that they can be stopped before the resulting error grows out of control.

Overall, we see that OMP is an extremely powerful, efficient, and robust algorithm that receives much less credit than it deserves. It is significantly faster than convex optimization techniques and is less sensitive to errors in sparsity level estimates.

## 6. Conclusion

Convex optimization has long been considered the gold standard compressive sensing recovery algorithm. Throughout the years, it has enjoyed significant theoretical development, putting it ahead of other faster algorithms, which up until recently, have been labeled as mere heuristics. The discovery of RIP-based performance guarantees for globalized matching pursuits such as CoSAMP and Iterative Thresholding has prompted a landslide of theoretical research into this class of algorithms. This paper presented near-optimal RIP-based guarantees for the more localized Orthogonal Matching Pursuit algorithm and the related method K-fold Orthogonal Matching Pursuit. In addition to deriving improved sufficient conditions guaranteeing the recoverability of strictly sparse signals, we also proved reconstruction error bounds for general signals possibly corrupted by measurement noise. While making significant contributions to OMP's theoretical development, we have failed to rigorously prove that OMP performs better than convex optimization, CoSAMP, Iterative Thresholding, etc. which do not suffer from the $\sqrt{T}$ blow-up factor that the latter algorithms successfully avoid. Thus, one may be led to believe that OMP is an inferior algorithm. Of course, the empirical evidence of Section 5 suggests otherwise. In practice, OMP and KOMP often outperform other algorithms in terms of accuracy, convergence, and stability. A possible explanation for this oxymoronic behavior is that RIP analysis considers worst case scenarios. In other words, it is possible to construct "bad" signals that convex optimization would recover more successfully than OMP. However, if an average case metric is used to theoretically evaluate the wide suite of compressive sensing algorithms, we are quite confident that OMP would rank very well.

## Acknowledgements

The author would like to thank Anna Gilbert and Martin Strauss from the University of Michigan for reviewing this work and providing comments and suggestions for improvement.